\documentstyle[eqsecnum,aps,prc,preprint,tighten]{revtex}

\begin{document}
\draft 
\preprint{TPR-96-19}

\title{Feynman rules in radial gauge}

\author{
S. Leupold}

\address{Institut f\"ur Theoretische Physik, Universit\"at 
        Regensburg,\\
        D-93040 Regensburg, Germany\\
        and \\
        Institut f\"ur Theoretische Physik, Justus-Liebig-Universit\"at 
        Giessen,\\
        D-35392 Giessen, Germany
}

\date{September 26, 1996}
\maketitle

\begin{abstract}
We present a complete set of Feynman rules for non-Abelian gauge fields
obeying the radial (Fock-Schwinger) gauge condition and prove the 
consistency with covariant gauge Feynman rules. 
\end{abstract}


\section{Introduction} 

Perturbation theory for non-Abelian gauge fields is usually formulated in 
covariant gauges. There the Feynman rules nearly can be read off from the 
momentum space representation of the Lagrange density including gauge fixing
and Faddeev-Popov ghost terms (c.f.~any textbook on gauge field theory, 
e.g.~\cite{pastar}). The only subtlety is the proper pole prescription for
the propagator $1/k^2$ which is achieved by the causal $\epsilon$ prescription,
i.e.~by the physically motivated requirement that particles (antiparticles)
propagate forward (backward) in time. 

Despite the great success of covariant
gauge Feynman rules there are situations where a different gauge choice is 
preferable. However it turned out during the last 15 years that things are 
much more involved if perturbation theory ought to be formulated in 
non-covariant gauges. The most prominent example is the case of temporal gauge
where an additional pole ($1/k_0^2$) appears in the gauge field propagator. 
Assuming the absence of ghost contributions and temporal propagator modes it 
was shown first in \cite{cara} that the actual value of gauge invariant 
quantities depends on the pole prescription for this additional propagator 
pole (see also \cite{Lan90} and references therein). According to \cite{leroy}
this pole is caused by the fact that the temporal gauge condition is 
insufficient to completely fix the gauge degrees of freedom. A complete gauge
fixing condition, on the other hand, automatically regulates this 
``gauge pole'', albeit breaks translational \cite{leroy} or
rotational \cite{palumbo} invariance, which at least makes the use of temporal
gauge Feynman rules less attractive. 

A second example is the Coulomb gauge
where singularities appear in the energy integrations of two loop diagrams
\cite{mitcheng,chengprl57,doust,taylor}. 
These singularities must be cured by a very subtle
prescription \cite{doust,taylor} or alternatively by the introduction of new
multi-gluon vertices \cite{christ}. As already pointed out in \cite{christ} 
these problems are due to operator ordering ambiguities which are difficult 
to handle in the path integral approach. 

These examples show that it is far from being trivial
to find consistent sets of Feynman rules for non-covariant gauges. With 
``consistent'' we simply mean that a given set of Feynman rules reproduces the 
same value for any gauge invariant quantity as the covariant gauge Feynman 
rules. 
The consistency of different sets of momentum space Feynman rules was 
systematically studied
by Cheng and Tsai \cite{mitcheng,chengprl57,chengwil}. A 
diagrammatic scheme was developed in \cite{mitcheng,chengwil} which 
serves to prove without analytic calculations that different sets of Feynman 
rules yield the same results for gauge invariant quantities. Cheng and Tsai 
called this the ``theorem of equivalence''. To apply this 
scheme to a given set of Feynman rules the propagators must not have any 
singularity, i.e.~every singularity must be regulated and the regulator must 
be kept finite during the course of calculation. E.g.~for the case of temporal
gauge it was shown \cite{chengprl57} that some prescriptions of the gauge 
pole mentioned above reintroduce ghost fields and/or temporal propagator modes 
which vanish if the regulator is taken to zero but nevertheless contribute to 
loop integrals as long as the regulator is kept finite. Hence the 
discrepancy between covariant and temporal gauge calculations observed in
\cite{cara} is due to the neglect of ghost and temporal propagator modes in the
temporal gauge using a specific gauge pole prescription, namely the principal
value prescription. 

In this article we will present a set of Feynman rules for the radial 
(Fock-Schwinger) gauge
and apply the diagrammatic scheme of Cheng and Tsai to this set of rules 
to show its
consistency with covariant gauge Feynman rules. The radial gauge with
its gauge condition 
\begin{equation}
x_\mu A^\mu(x) = 0  \label{radgaugecond} 
\end{equation}
was introduced long time ago \cite{fock,schwing} and rediscovered several 
times (e.g.~\cite{cron}). It found widespread use in the context of QCD 
sum-rules (e.g.~\cite{shif}) since it enormously simplifies the task of 
organizing the operator
product expansion of QCD $n$-point functions in terms of gauge invariant
quantities by expressing the gauge potential via the gauge covariant
field strength tensor. 

In (\ref{radgaugecond}) the origin is singled out by the gauge condition. 
Therefore this gauge condition breaks translational invariance and 
perturbation theory cannot be formulated in
momentum space as usual but must be set up in coordinate space. For this reason
only few efforts have been made to establish perturbation theory for
radial gauge \cite{kumm,moda,menot,menot2,delbourgo,leuwei96}. 
Most of the discussion was concentrated on
the question about the properties of the free radial gauge propagator while
less efforts where made to explore the other Feynman rules, especially to 
figure out if ghost fields truly decouple. 
Recently a radial propagator with surprising properties was proposed in 
\cite{leuwei96}. 

In the next Section we briefly recapitulate the results found in 
\cite{leuwei96}. In Sec.~\ref{sec:toe} we generalize the diagrammatic
scheme of Cheng and Tsai, originally developed for momentum space,  to 
coordinate space Feynman rules. In Sec.~\ref{sec:radfeynrules} we apply this
scheme to a complete set of radial gauge Feynman rules including the radial 
gauge field propagator proposed in \cite{leuwei96}. An additional check for
the consistency of this set is presented in Sec.~\ref{sec:slavtay} using 
the Slavnov-Taylor identities of radial gauge. Finally we summarize our
results in Sec.~\ref{sec:sum}.

\section{The radial gauge propagator} \label{sec:review} 

In the following we work in a $D$-dimensional Minkowski space, i.e.~with one
time and $D-1$ space dimensions. 

In \cite{leuwei96} it was shown that the (full) radial gauge field propagator
can be expressed in terms of a gauge invariant Wilson loop in the following
way:
\begin{equation}
D^{ab}_{\mu\nu}(x,y):=\langle A^a_\mu(x) A^b_\nu(y) \rangle
= \delta^{ab} {2N\over N^2 -1} {1\over (ig)^2} 
\lim_{x'\to x \atop   y'\to y}
\partial_\mu^x \partial_\nu^y \, W_1(x,x',y,y') \,.  \label{propwil}
\end{equation}
where $N$ is the numbers of colors and the Wilson loop $W_1$ connects the
sequence of points $0,y',y,0,x',x,0$ along straight lines 
(c.f.~Fig.~\ref{faecher}). As a Wilson loop with cusps and self-intersections
$W_1$ has so-called ``cusp singularities'' (c.f.~\cite{brandt} and references
therein) appearing at quadratic (and higher) order 
in the coupling constant. Due to the prefactor $1/g^2$ in (\ref{propwil}) 
these singularities show up at order $g^0$, i.e.~already for the free radial
propagator. The explicit form of the latter is given by
\begin{eqnarray}
  \lefteqn{\langle A^a_\mu(x) A^b_\nu(y) \rangle_0 =} 
\nonumber\\&&  = {\Gamma(D/2-1) \over
    4\pi^{D/2}} \,\delta^{ab} \, \bigg( 
g_{\mu\nu} [(x-y)^2-i\epsilon]^{1-D/2}
  \nonumber\\ && \phantom{mmm} -\partial_\mu^x \int\limits^1_0 \!\! ds
  \, x_\nu\,[(sx-y)^2-i\epsilon]^{1-D/2} 
-\partial_\nu^y \int\limits^1_0 \!\! dt
  \, y_\mu \,[(x-ty)^2-i\epsilon]^{1-D/2} \nonumber\\ && \phantom{mmm}
  +\partial_\mu^x \partial_\nu^y \underbrace{\int\limits^1_0 \!\! ds
    \int\limits^1_0 \!\! dt\, x\cdot y \,
    [(sx-ty)^2-i\epsilon]^{1-D/2}}_{\textstyle \sim {\textstyle 1\over
      \textstyle 4-D}} \bigg) \,.
\label{freeradprop}
\end{eqnarray}
which is obviously singular for $D \to 4$. That the 
free radial propagator diverges in four dimensional space was already pointed
out earlier in \cite{moda}. The reason for this singularity was presented in
\cite{leuwei96}. It is a surprising result that the renormalization
properties of a special class of Wilson loops influence even the free radial
propagator. 
However it was argued in \cite{leuwei96} that this singularity is even 
necessary to reproduce correct results for gauge invariant quantities 
calculated in radial gauge. 

Note that we have used dimensional regularization for convenience. 
Of course the cusp singularities of Wilson loops and therefore also the 
singularity of the radial propagator might be regularized in a different way 
as well. In any case the regulator has to be kept finite during the 
calculation which however is required for loop calculations anyway. In 
addition a finite regulator is necessary to make the diagrammatic scheme of 
Cheng and Tsai applicable to radial gauge Feynman rules.

\section{The theorem of equivalence in coordinate space} \label{sec:toe}

The theorem of equivalence was originally formulated by Cheng and Tsai for 
momentum space Feynman rules \cite{mitcheng}. In this section we generalize 
it to coordinate space. The basic statement is that all gauge invariant 
quantities are independent of the arbitrary function $\Delta_\mu(x,y)$ if the
following Feynman rules are used:\\
The free gauge field propagator 
\begin{equation}
D^{ab}_{\mu\nu}(x,y) = \delta^{ab} \left[
D^F_{\mu\nu}(x,y) - \partial_\mu^x\,\Delta_\nu(x,y) 
- \partial_\nu^y \, \Delta_\mu(y,x) \right]  \,,  \label{belproports}
\end{equation}
the ghost field 
\begin{equation}
G^{abc}_\mu(y,x) =  -i g f^{abc} \left[
(g_{\mu\nu} \Box_x - \partial^x_\mu \partial^x_\nu )
\Delta^\nu(y,x) + \partial^\nu_y\,D^F_{\mu\nu}(x,y)
\right]  \,,
\label{belgeistorts}
\end{equation}
and the common three and four gluon vertices. Here $D^F_{\mu\nu}$ denotes the
Feynman propagator given in Appendix \ref{anh:feyn}, the ``ghost field'' is 
the product of ghost-gluon vertex and ghost propagator, and the coordinate 
space representation of the three and four gluon vertices is also given in
Appendix \ref{anh:feyn}. Note that we get the expressions for Feynman gauge
for vanishing $\Delta_\mu$. Hence the ghost field (\ref{belgeistorts}) can be
written as 
\begin{equation}
G^{abc}_\mu(y,x) =  -i g f^{abc} 
(g_{\mu\nu} \Box_x - \partial^x_\mu \partial^x_\nu )
\Delta^\nu(y,x) + G^{F\,abc}_\mu(y,x) \,.
\label{ctwla}
\end{equation}

To prove that all gauge invariant quantities are independent of $\Delta_\mu$ 
we observe that in the expression for the gauge propagator (\ref{belproports})
$\Delta_\mu$ is always accompanied by a partial derivative. This
reflects the fact that the gauge modes and the longitudinal modes coincide in 
the free field case of a gauge theory. By partial integration we arrive at an
expression where this partial derivative acts on a vertex which is connected 
to the propagator. The partial derivative acts on a three gluon vertex 
(\ref{dreigluon}) as 
\begin{eqnarray}
\partial^\lambda_w \, T_{\lambda \mu \nu}^{abc}(w,x,y) &=&
ig f^{abc} \left[
\delta(w-y) (g_{\nu\mu} \Box_w - \partial^w_\nu \partial^w_\mu ) \delta(w-x)
\right.\nonumber\\ && \left.\phantom{mmm}
-\delta(w-x) (g_{\mu\nu} \Box_w - \partial^w_\mu \partial^w_\nu ) \delta(w-y)
\right]  \label{ctwlb}  \,.
\end{eqnarray}
Obviously the operators inside the brackets project on transverse components,
i.e. 
\begin{equation}
(g_{\nu\mu} \Box_w - \partial^w_\nu \partial^w_\mu ) \partial^\mu_w = 0 \,.
\label{ctwlbtrans}
\end{equation}
These operators act on the next propagator which is connected to the vertex 
at $x,\mu,b$ or $y,\nu,c$, respectively, e.g.  
\begin{eqnarray}
\lefteqn{
ig f^{abc} \int\!\!d^D\!x\, 
(g_{\nu\mu} \Box_w - \partial^w_\nu \partial^w_\mu ) \delta(w-x)
D^{\mu\mu'}_{bb'}(x,x')
}\nonumber\\ &&
= -g f^{ab'c} g_{\nu}^{\phantom{\nu}\mu'} \delta(w-x') 
+ \partial^{\mu'}_{x'}\, G^{ab'c}_\nu(x',w)  \,. \label{ctwlc}
\end{eqnarray}
This yields ghost type structures as well as contractions 
$-g f_{ab'c} g_{\nu}^{\phantom{\nu}\mu'} \delta(w-x')$. As we shall show in the
following the former cancel gauge dependent contributions from ghost loops
while the latter are related to four gluon vertices. 
From (\ref{ctwlb}) and (\ref{ctwlc}) we get 
\begin{eqnarray}
\lefteqn{\int\!\! d^D\!x\,d^D\,y\,\partial^\lambda_w \, 
T_{\lambda \mu \nu}^{abc}(w,x,y) 
D^{\mu\mu'}_{bb'}(x,x') D^{\nu\nu'}_{cc'}(y,y') = } \nonumber \\ 
&& \!\!\!\! \int\!\! d^D\!y \left[ 
-g f^{ab'c} g_{\nu}^{\phantom{\nu}\mu'} \delta(w-x') \,\delta(w-y)
+ \partial^{\mu'}_{x'}\, G^{ab'c}_\nu(x',w) \,\delta(w-y) 
\right] D^{\nu\nu'}_{cc'}(y,y')
\nonumber \\ && \!\!\!\!{}
- \int\!\! d^D\!x \left[ 
-g f^{abc'} g_{\mu}^{\phantom{\mu}\nu'} \delta(w-y') \,\delta(w-x)
+ \partial^{\nu'}_{y'}\, G^{abc'}_\nu(y',w) \,\delta(w-x) 
\right] D^{\mu\mu'}_{bb'}(x,x')
\,. \label{neuein}\end{eqnarray}
So far this looks very complicated. However a graphical representation of this
formula is rather simple and indeed all calculations which are necessary
to prove the theorem of equivalence can be performed graphically. Once all 
rules are stated, there is no need to write down any analytical expression. 
Formula (\ref{neuein}) is depicted in Fig.~\ref{pardrei}. In the following 
all lines 
labeled with $D$ denote gluon propagators. Three and four gluon vertices are
dotted. $G$ denotes a ghost field which includes a vertex and a propagator 
line according to the definition (\ref{belgeistorts}). An arrow at the end of 
a propagator line denotes a partial derivative acting on the respective 
vertex. Finally an arrow appearing as a label denotes the contraction mentioned
above, i.e.~the first contribution on the r.h.s.~of (\ref{ctwlc}). 

Instead of presenting a complete set of diagrammatic rules first and applying
it to an example afterwards we will choose a simple example right now and 
introduce the necessary rules step by step as soon as we need them. As an 
example we take the two loop vacuum contributions shown in 
Fig.~\ref{bubbleall}. The combinatorial factors for these diagrams are $1/12$
for the diagram involving three gluon vertices, $1/8$ for the four gluon vertex
diagram and $1/2$ for the diagram with a ghost loop. It is useful to state the
relative statistical weights and signs of all diagrams explicitly. This is 
depicted in Fig.~\ref{bubblesym}. An overall factor $1/12$ is suppressed. In
addition the diagrams are symmetrized with respect to the propagator lines. 

To prove the gauge invariance of the sum shown in Fig.~\ref{bubblesym} we 
have to show that the result remains the same if all $D$'s and $G$'s are 
replaced by $D^F$ and $G^F$, respectively. To do that we will even prove a 
somewhat more rigorous statement: The sum remains the same if all fields 
$D$ and $G$ labeled by the same number are changed to $D^F$ and $G^F$ no 
matter of the gauge of the fields with different numbers. Thus we will prove
in the following that without changing the actual value of the whole sum 
$D_1$ and $G_1$ can be replaced by $D^F_1$ and $G^F_1$ without any reference 
to the explicit form of $D_{2,3}$ and $G_{2,3}$. This is enough to prove
gauge invariance since this scheme can be applied to the diagrams of 
Fig.~\ref{bubblesym} consecutively for three times: In the first step $D_1$
and $G_1$ are changed to $D^F_1$ and $G^F_1$. Next $D_2$ and $G_2$ can be 
replaced by $D^F_2$ and $G^F_2$ in the same way now dealing with $D^F_1$, 
$G^F_1$, $D_3$, and $G_3$. Finally $D_3$ and $G_3$ are replaced by their 
Feynman gauge expressions starting from diagrams where the lines labeled by
1 and 2 already are in Feynman gauge. Thus we will prove that the gauge of 
{\it one} line in {\it all} diagrams can be changed without changing the value
of the sum of diagrams. 

We start with the first diagram of Fig.~\ref{bubblesym} and decompose $D_1$ 
in its Feynman and $\Delta$ parts according to (\ref{belproports}). This is
shown in Fig.~\ref{propzer} where a partial integration has already been 
performed changing the relative sign of the $\Delta$ contributions with respect
to the Feynman part. 

Now we use the diagrammatic rule of Fig.~\ref{pardrei} to obtain 
Fig.~\ref{erster}. 

In this way we get two types of diagrams. The first type involves ghosts and
partial derivatives (arrows). The latter act on the second three gluon vertex 
and we can use (\ref{ctwlb}) again. In addition to (\ref{ctwlc}) we need a 
relation where the propagator $D$ is replaced by $\Delta$. However this is
nothing but (\ref{ctwla}). Using (\ref{ctwlb}), (\ref{ctwlc}) and (\ref{ctwla})
we get 
\begin{eqnarray}
\lefteqn{\int\!\! d^D\!x\,d^D\!y\,\partial^\lambda_w \, 
T_{\lambda \mu \nu}^{abc}(w,x,y) 
\Delta^\mu(x',x) D^{\nu\nu'}_{cc'}(y,y') = } \nonumber \\ &&
\int\!\! d^D\!y \left[ 
G^{F\,abc}_\nu(x',w)\,\delta(w-y) - G^{abc}_\nu(x',w)\,\delta(w-y) 
\right] D^{\nu\nu'}_{cc'}(y,y')
\nonumber \\ && {}
- \int\!\! d^D\!x \left[ 
-g f^{abc'} g_{\mu}^{\phantom{\mu}\nu'} \delta(w-y') \,\delta(w-x)
+ \partial^{\nu'}_{y'}\, G^{abc'}_\mu(y',w) \,\delta(w-x) 
\right] \Delta^\mu(x',x)  \label{neuein2} 
\end{eqnarray}
which is shown in Fig.~\ref{delstd}. 

For a diagram of Fig.~\ref{erster} with a ghost line we get Fig.~\ref{zweiter}.
As a first example how the cancelation of gauge dependent diagrams shows up
we note that the second diagram on the r.h.s.~of Fig.~\ref{zweiter} is 
identical to diagram G of Fig.~\ref{bubblesym}. These diagrams cancel each 
other (note the minus sign in front of the third diagram of 
Fig.~\ref{erster}). 

The last diagram of Fig.~\ref{zweiter} can be transformed in a diagram with a 
ghost loop in the following way: The action of the arrow (partial derivative)
on the ghost vertex can be split up in a sum of two terms using the product 
rule of differentiation.\footnote{It is interesting to mention that the 
momentum space equivalent of this product rule is the energy-momentum 
conservation at the vertex.} On the one hand the partial derivative acts on a 
ghost field as (cf.~(\ref{ctwla}))
\begin{equation}
\partial_\mu^x G_{abc}^\mu(y,x) = \partial_\mu^x G_{F\,abc}^\mu(y,x) 
= -g f_{abc} \delta(x-y)  \,,
\end{equation}
and on the other hand it acts on the upper line. This is shown in 
Fig.~\ref{einser}. The corresponding equation is 
\begin{eqnarray}
\lefteqn{\int\!\! d^D\!x \,\delta(v-x) \,\partial_\mu^x 
\left[ G^\mu_{abc}(y,x) \,\delta(x-w) \right] =} \nonumber \\ &&  {}
- \int\!\! d^D\!x \,\delta(x-w) \,\partial_\mu^x
\left[ G^\mu_{abc}(y,x) \,\delta(v-x) \right]
- \int\!\! d^D\!x \, \delta(v-x) \, g f_{abc} \delta(x-y)\,\delta(x-w) 
\end{eqnarray}
and the symbol ${\bf 1}$ in Fig.~\ref{einser} denotes the last integrand. 

In this way the last diagram of Fig.~\ref{zweiter} yields two new diagrams as
shown in Fig.~\ref{pull}. Since the only difference between ${\bf 1}$ and the 
contraction arrow is a $g_{\mu\nu}$-term it is easy to get the last line of 
Fig.~\ref{pull}. 

Now we collect all diagrams which can be deduced from the two diagrams 
involving $G$'s shown in Fig.~\ref{erster} and compare the result, 
Fig.~\ref{vierter}, with the original diagrams of Fig.~\ref{bubblesym}:
In addition to the ten diagrams of Fig.~\ref{vierter} we get ten analogous 
diagrams not shown here from the last diagram of Fig.~\ref{propzer}. We refer
to these diagrams as the ``reflected'' diagrams of the ones shown in 
Fig.~\ref{vierter}. Diagrams
a and b of Fig.~\ref{vierter} replace $G_1$ by $G^F_1$ in 
Fig.~\ref{bubblesym}G. The same is achieved by f and g in 
Fig.~\ref{bubblesym}I. The corresponding reflected diagrams 
transform Fig.~\ref{bubblesym}F and J in the same way. 
Diagrams c and j as well as e and h cancel each other. Diagram d and the 
reflected diagram of i replace $D_1$ by $D_1^F$ in 
Fig.~\ref{bubblesym}H. In the same way Fig.~\ref{bubblesym}E is
transformed by Fig.~\ref{vierter}i and the reflected diagram of 
Fig.~\ref{vierter}d. 

Thus for the diagrams E-J of Fig.~\ref{bubblesym} we have already succeeded
in replacing $D_1$ and $G_1$ by $D^F_1$ and $G^F_1$. We are left with the
four gluon vertex diagrams B and C on the one hand\footnote{Diagram D remains
unchanged since no $D_1$ appears there.} and with the ghost-free
diagrams of Fig.~\ref{erster} and their reflected diagrams on the other hand. 
In Fig.~\ref{vierzer} we have decomposed $D_1$ of Fig.~\ref{bubblesym}B  
using (\ref{belproports}). 

Obviously we need a relation which connects contraction arrows with four gluon 
vertices. Indeed the following relation holds 
\begin{eqnarray}
\lefteqn{-\int\!\! d^D\!w' \, 
g f_{abe} g^{\lambda}_{\phantom{\lambda}\lambda'} \delta(v-w)\,\delta(v-w')\,
T_{ecd}^{\lambda'\mu\nu}(w',x,y) } \nonumber \\ && {}
+ \int\!\! d^D\!x' \,
g f_{aec} g^{\mu}_{\phantom{\mu}\mu'} \delta(v-x) \,\delta(v-x')\,
T_{edb}^{\mu'\nu\lambda}(x',y,w)
\nonumber \\ && {}
+ \int\!\! d^D\!y' \,
g f_{aed} g^{\nu}_{\phantom{\nu}\nu'} \delta(v-y) \,\delta(v-y')\,
T_{ebc}^{\nu'\lambda\mu}(y',w,x)
= \partial_\kappa^v Q^{\kappa\lambda\mu\nu}_{abcd}(v,w,x,y)   \,. 
\end{eqnarray}
This is shown in Fig.~\ref{pfeilvier}. Connecting the upper legs with each
other and also the lower legs we find that the last but one diagram of 
Fig.~\ref{erster} and its reflected diagram exactly cancel the last diagram
of Fig.~\ref{vierzer}.\footnote{Note that the third diagram of 
Fig.~\protect\ref{pfeilvier} vanishes due to its color structure if the upper
legs are connected.}  The same
holds true for the diagrams where $D_2$ is replaced by $D_3$, 
i.e.~Fig.~\ref{bubblesym}C instead of B and the second diagram of 
Fig.~\ref{erster} instead of the fourth. 

We conclude that the last two diagrams of Fig.~\ref{propzer} can be decomposed
in a way that the resulting diagrams can be used to replace 
$D_1$ and $G_1$ by $D^F_1$ and $G^F_1$ in Fig.~\ref{bubblesym}B-J. This proves
the theorem of equivalence for our example of two loop vacuum bubbles. 

In the same way the theorem of equivalence can be proven for higher loop vacuum
diagrams, scattering amplitudes and Wilson loops. There are a few more 
diagrammatic rules collected in Appendix \ref{rechen} which we did not need 
here for our explicit example. However they become important for higher order
Feynman diagrams. The complete set of diagrammatic rules, Figs.~\ref{pardrei},
\ref{delstd}, \ref{einser}, \ref{pfeilvier}, \ref{nurvier}, \ref{pfeileins},
and \ref{wilo}, can be applied to any gauge invariant quantity in any order of
the coupling constant. The necessary steps to prove the gauge invariance of
a sum of loop diagrams are as follows: 

\begin{itemize}
 \item Symmetrize all diagrams with respect to all lines (as shown in 
       Fig.~\ref{bubblesym}).
 \item Start with the diagrams which have only three-gluon vertices. 
       Decompose one gluon propagator according to (\ref{belproports}).
 \item Use Fig.~\ref{pardrei} and \ref{delstd} as long 
       as there are $\partial_\mu$-arrows pointing at a three-gluon vertex.
 \item Use Fig.~\ref{einser} to pull arrows out of closed ghost loops
       (c.f.~Fig.~\ref{pull}).
 \item At this stage all $\Delta_\mu$-dependent diagrams which have
       neither contraction arrows nor four-point vertices should drop out.
 \item Finally, diagrams with different topology are related by the rules
       of Figs.~\ref{pfeilvier}, \ref{nurvier}, \ref{pfeileins}. To get an 
       idea which diagrams are related 
       it is instructive to contract the lines with contraction arrows.
 \item For Wilson loops the additional relation shown in Fig.~\ref{wilo} has
       to be taken into account. 
\end{itemize}

This scheme can also be generalized to cases
where the Yang-Mills fields are coupled to matter fields. In principle for
each new vertex two new diagrammatic rules must be found, which have to 
show how the
two introduced arrows (denoting the partial derivative and the contraction) act
on this new vertex. For the case of full QCD (albeit using momentum space 
Feynman rules) this complete set of diagrammatic rules is given in 
\cite{achh96} and applied to three loop diagrams. 
Independently of the original work of Cheng and Tsai \cite{mitcheng}  
a similar set of diagrammatic rules was developed recently by Feng and Lam 
also for momentum space Feynman rules and applied to QCD \cite{feng96} and 
electro-weak theory \cite{feng96a} to simplify perturbative calculations by 
choosing an appropriate gauge. 
 
With the diagrammatic scheme developed in this Section we return to the case 
of radial gauge Feynman rules.

\section{Application to radial gauge} \label{sec:radfeynrules}

Now we have developed all tools necessary to present a consistent set of 
radial gauge Feynman rules. The free radial gauge field propagator is given
in (\ref{freeradprop}). To make the scheme of Cheng and Tsai applicable the
regulator (here $D-4$) has to be kept finite until the end of the calculation. 
Note however that the regularized propagator already fulfills the radial gauge
condition. This is in contrast to the case of temporal gauge. There the 
temporal mode of the propagator vanishes only if the regulator vanishes. This
mode, however, couples to the ghost fields. As we shall see in the following 
ghost fields decouple for radial gauge since already the regularized propagator
strictly satisfies the gauge condition. 

Comparing (\ref{belproports}) with (\ref{freeradprop}) we get 
\begin{eqnarray}
\Delta_\nu(x,y) = x^\alpha \int\limits_0^1 \!\! ds\, D^F_{\alpha\nu}(sx,y)
- {1\over 2} \partial^\alpha_y 
\int\limits_0^1 \!\! ds \int\limits_0^1 \!\! dt \, x\cdot y\,
D^F_{\alpha\nu}(sx,ty) \,.
\end{eqnarray}
After some simple algebraic manipulations we find that according to 
(\ref{belgeistorts}) the appropriate ghost field is given by 
\begin{eqnarray}
G_\nu^{abc}(x,y) = -ig f^{abc} \left[
x_\nu \int\limits_0^1\!\! ds\, \delta(sx-y) 
- \partial^\alpha_y \, D^F_{\alpha\nu}(0,y)  
\right]  \,.  \label{radgeistent}
\end{eqnarray}
It is easy to see that 
\begin{eqnarray}
G_\nu^{abc}(x,y) \sim y_\nu 
\end{eqnarray}
holds. On account of the gauge condition (\ref{radgaugecond}) we find
\begin{eqnarray}
G_\nu^{abc}(x,y) D^{\nu\nu'}_{cc'}(y,y') = 0 \,. 
\end{eqnarray}
Thus the ghost fields decouple from the gauge fields. Of course one
expects that to happen for gauge fields obeying an algebraic gauge condition.
As mentioned in the introduction this is however a non-trivial problem for 
the case of temporal gauge. This example teaches us that it is not sufficient
to simply study the Faddeev-Popov ghost determinant. A more 
careful examination, e.g.~like the one presented here, is necessary to state 
the decoupling of ghost fields. 

One subtlety has to be clarified before we can apply the scheme developed in
the last Section to our case at hand: One frequently used tool for the 
manipulations of Feynman diagrams was the partial integration of derivatives 
$\partial_\mu^x$. For all the statements of the last Section to be valid
we have to make sure that this partial integration produces no surface terms.
Clearly this depends on the actual form of $\Delta$. Contrary to other radial
gauge propagators presented earlier the propagator (\ref{freeradprop}) is 
well-defined at the origin (for a detailed discussion of this point 
c.f.~\cite{leuwei96}). Thus the behavior at the origin cannot be a source of 
surface terms. However things are more complicated at infinity. A careful
inspection of the propagator (\ref{freeradprop}) shows that for large $x$ 
the first and third contribution decreases like $1/(x^2)^{D/2-1}$. Also 
the propagator for the Feynman gauge (\ref{Feynprop}) shows this rate of 
decrease and there no problems with surface terms appear. However
the second and fourth contribution of (\ref{freeradprop}) drops only 
like $1/(x^2)^{(D-3)/2}$. This weaker rate of decrease is always accompanied
by a partial derivative $\partial_\mu^x$. The worst case that might appear 
is that all legs at a vertex are of this type. A typical example for such a 
coincidence is 
\begin{eqnarray}
\int\!\! d^D\! w\, \int\!\! d^D\! x \, \int\!\! d^D\! y \,
T_{abc}^{\lambda\mu\nu}(w,x,y) \, 
\partial_\lambda^w \Delta_{\lambda'}(w,w')
\partial_\mu^x \Delta_{\mu'}(x,x')
\partial_\nu^y \Delta_{\nu'}(y,y')  \,.  \label{wouldbeprob} 
\end{eqnarray}
Inserting (\ref{dreigluon}) and evaluating the $\delta$-functions 
we are left with a $D$ dimensional integral and an integrand with a partial 
derivative and three $\partial\Delta$ contributions. Hence the behavior for
large $x$ is given by $\int\!\!dx/(x^2)^{D-7/2}$ which might cause a 
logarithmic surface term for $D \to 4$. Before discussing this case in more
detail we note that this actually is the only problematic case. 
If any of the $\partial\Delta$ contributions is replaced by a term which 
falls off like $1/(x^2)^{D/2-1}$ no surface term shows up if $D$ is in the
vicinity of 4. If there is any surface term it would be caused by a partial
integration in (\ref{wouldbeprob}). Fortunately we can calculate this 
potential surface term explicitly by inserting the definition for the 
three gluon vertex (\ref{dreigluon}) and using (\ref{ctwlb}). We find 
\begin{eqnarray}
\lefteqn{\int\!\! d^D\! w\, \int\!\! d^D\! x \, \int\!\! d^D\! y \,
T_{abc}^{\lambda\mu\nu}(w,x,y) \, 
\partial_\lambda^w \Delta_{\lambda'}(w,w')
\partial_\mu^x \Delta_{\mu'}(x,x')
\partial_\nu^y \Delta_{\nu'}(y,y')} 
\nonumber \\ &&
+ \int\!\! d^D\! w\, \int\!\! d^D\! x \, \int\!\! d^D\! y \,
\partial_\lambda^w T_{abc}^{\lambda\mu\nu}(w,x,y) \, 
\Delta_{\lambda'}(w,w')
\partial_\mu^x \Delta_{\mu'}(x,x')
\partial_\nu^y \Delta_{\nu'}(y,y')
= 0 \,.  
\end{eqnarray}
Thus also in this problematic case no surface term appears. 

We conclude that for the manipulations of radial gauge Feynman diagrams 
partial integrations can be performed without any problems. Therefore 
the diagrammatic scheme presented in the last Section can be applied to the
radial gauge Feynman rules. This proves that a consistent set of Feynman 
rules for the radial gauge is given by the radial propagator 
(\ref{freeradprop}) and the usual three and four gluon vertices 
(\ref{dreigluon}) and (\ref{viergluon}), respectively. 
The ghost fields decouple (for the temporal
gauge this depends explicitly on the pole prescription) and no additional 
vertices are necessary (for the Coulomb gauge one has to encounter that 
problem).

\section{Slavnov-Taylor identities} \label{sec:slavtay} 

Slavnov-Taylor identities for Yang-Mills fields formulated in radial gauge 
have already been
derived in \cite{kumm}. Besides the diagrammatic scheme presented in 
Sec.~\ref{sec:toe} these identities also serve to check the consistency of a 
set of Feynman rules. In addition these identities are an important ingredient 
to set up a renormalization scheme for radial gauge perturbation theory. For
this latter aspect we refer to the remarks made in \cite{kumm}. 

In the following we only study the consequences of the Slavnov-Taylor 
identities for the gluon self energy. 
In principle also higher one-particle irreducible
$n$-point functions can be studied. 
We start with the generating functional for Yang-Mills fields in radial gauge  
and introduce a Lagrange multiplier field $C$:
\begin{eqnarray}
W[J]&:=&
\int\!\! {\cal D}A \, \,
\delta\!\left( x_\mu A^\mu(x) \right)
\nonumber\\ && 
\times \exp\left(i\int\!\! d^D\!x \,( {\cal L}_{\rm YM}[A](x) 
+ J^a_\mu(x) A_a^\mu(x) ) \right)
\nonumber\\  
&=&  \int\!\! {\cal D}[A,C] \, 
\exp\left(i\int\!\! d^D\!x \, ( {\cal L}_{\rm YM}[A](x) 
+ C^a(x) x_\mu A_a^\mu(x) 
+ J^a_\mu(x) A_a^\mu(x)) \right) 
\,.
\end{eqnarray}
In addition we introduce 
the generating functional for vanishing source terms, 
\begin{equation}
W_0 := W[J=0]   \,, 
\end{equation}
and full and free Green's functions for an arbitrary quantity $\xi$:
\begin{equation}
\langle \xi \rangle
:= {1\over W_0} \int\!\! {\cal D}[A,C] \, \xi
\exp (i S [A,C])  
\end{equation} 
and 
\begin{equation}
\langle \xi \rangle_0 := 
\langle \xi \rangle \left.\vphantom{\int} \right\vert_{g=0}  \,, 
\end{equation}
respectively. Here the action $S$ is defined as 
\begin{eqnarray}
S[A,C] := 
\int\!\! d^D\!x \, ( {\cal L}_{\rm YM}[A](x) 
+ C^a(x) x_\mu A_a^\mu(x)) \,. 
\end{eqnarray}

It is well-known that the operator 
\begin{equation}
{\cal O}^{\mu\nu}(x) = g^{\mu\nu} \Box_x - \partial_x^\mu \partial_x^\nu
\label{deffreeop} \end{equation}
which appears in the free part of ${\cal L}_{\rm QCD}$ 
is not invertible due to the gauge invariance of the theory. However in the 
enlarged space of $A$ and $C$ fields the operator ${\cal O}$ can be inverted. 
We denote propagator and self energy in the larger space with 
\begin{eqnarray}
\lefteqn{\bar D^{ab}_{\Theta\Phi}(x,y) := \left(
\begin{array}{cc}
\langle A^a_\mu(x) \, A^b_\nu(y) \rangle   & 
\langle A^a_\mu(x) \, C^b(y) \rangle   \\
\langle C^a(x) \, A^b_\nu(y) \rangle      & 
\langle C^a(x) \, C^b(y) \rangle  
\end{array}
\right)} \nonumber \\ &&
 = \left(
\begin{array}{cc}
D_{\mu\nu}^{ab}(x,y)  &
\langle A^a_\mu(x) \, C^b(y) \rangle   \\
\langle C^a(x) \, A^b_\nu(y) \rangle      & 
\langle C^a(x) \, C^b(y) \rangle  
\end{array}
\right)
\end{eqnarray}
and $\bar \Pi^{ab}_{\Theta\Phi}(x,y)$, respectively. The large Greek letters
denote numbers from 0 to 4. For 0 to 3 they coincide with the Lorentz indices 
of the gauge fields while 4 denotes the additional Lagrange multiplier field 
$C$. 

The Dyson-Schwinger equation turns out to be 
\begin{equation}
\bar D = \bar D_0 + \bar D_0 \bar\Pi \bar D  \,,  \label{dysch}
\end{equation}
where ``multiplication'' includes summation over all indices and integration 
over the space time variables, i.e.~$\bar\Pi \bar D $ is an abbreviation for
\begin{equation}
\sum\limits_{\Xi = 0}^4 \sum\limits_{b=1}^{N^2-1} \int\!\! d^D\!x \,
\bar \Pi_{\Theta\Xi}^{ab}(w,x)\,  \bar D^{\Xi\Phi}_{bc}(x,y)  \,.
\end{equation}
$\bar D_0$ is the free contribution of $\bar D$. The inverse is the operator
which appears in the free action (cf.~(\ref{deffreeop})):
\begin{equation}
\left( \bar D_0^{-1} \right)_{\Theta\Phi}^{ab}(x,y) = -i \delta^{ab}
\left(
\begin{array}{cc}
{\cal O}_{\mu\nu}(x)  & x_\mu \\
x_\nu & 0 
\end{array}
\right) \delta(x-y)  \,.
\end{equation}
The Dyson-Schwinger equation (\ref{dysch}) can be transformed to 
\begin{equation}
\bar \Pi = \bar D_0^{-1} \,(\bar D - \bar D_0 ) \bar D^{-1}  \,,
\end{equation}
which can be further evaluated using the Slavnov-Taylor identities 
\begin{equation}
\langle A^a_\mu(x) \, C^b(y) \rangle  = 
\langle A^a_\mu(x) \, C^b(y) \rangle_0 
\end{equation}
and 
\begin{equation}
\langle C^a(x) \, C^b(y) \rangle  = 0 
\end{equation}
stated in \cite{kumm}. We find 
\begin{equation}
\bar D - \bar D_0 = 
\left(
\begin{array}{cc}
D - D_0 & 0 \\
0 & 0
\end{array}
\right)  \,;
\end{equation}
hence:
\begin{eqnarray}
\bar \Pi 
&=& -i \left( \begin{array}{cc} {\cal O} & x \\ x & 0 \end{array} \right)
\left( \begin{array}{cc} D - D_0 & 0 \\ 0 & 0 \end{array} \right)
\bar D^{-1}
\nonumber \\ && \nonumber \\
&=& -i \left( 
\begin{array}{cc} {\cal O} (D - D_0) & 0 \\ x (D - D_0) & 0 \end{array}
\right)
\bar D^{-1} 
= -i \left( 
\begin{array}{cc} {\cal O} (D - D_0) & 0 \\ 0 & 0 \end{array}
\right)
\bar D^{-1}
\nonumber \\ && \nonumber \\
&=& -i \left(
\begin{array}{cc} * & * \\ 0 & 0 \end{array}
\right)  \,. \label{self0}
\end{eqnarray}
where we have used the gauge condition (\ref{radgaugecond}) in the second line.
Stars denote arbitrary values. Since the self energy is symmetric with respect
to an exchange of all variables we conclude that it has a non-vanishing 
contribution in the gauge sector, only: 
\begin{equation}
\bar \Pi = \left(
\begin{array}{cc}
\Pi & 0 \\ 0 & 0
\end{array}
\right)  \,.
\end{equation}
In addition we get the transversality of the self energy by multiplying 
(\ref{self0}) with $(\partial_\mu ,0)$ from the left
\begin{equation}
  (\partial \Pi ,0) =  (\partial , 0) (-i) \left(
    \begin{array}{cc}
      {\cal O} (D - D_0) & 0 \\
      0 & 0
    \end{array}
  \right) \bar D^{-1}
  = (0,0)  \,\,,
\end{equation}
due to $\partial {\cal O} =0$ (cf.~(\ref{deffreeop})). 

We conclude that as a consequence of the Slavnov-Taylor identities the radial
gauge gluon self energy obeys the transversality condition 
\begin{equation}
\partial^x_\mu \Pi^{\mu\nu}(x,y) = 0  \label{tracon}
\end{equation} 
We will finish this Section by checking this equation 
for the lowest order contribution in the coupling constant. Fortunately we 
can use the diagrammatic rules of Sec.~\ref{sec:toe}. 
Since the ghost fields decouple only
two diagrams contribute to the self energy at order $g^2$. They are shown in
Fig.~\ref{trglse}. To derive the r.h.s.~we have made use of the fact that the
first diagram of Fig.~\ref{trglse} is symmetric with respect to an exchange 
of upper and lower line. In addition we have applied the diagrammatic rule of 
Fig.~\ref{pardrei}. Using the relation shown in Fig.~\ref{pfeilvier} we find
that the last diagram in Fig.~\ref{trglse} 
is exactly canceled by the last but one diagram. 
The remaining diagram involving $G$ vanishes since $G$ is contracted with
a radial propagator $D$, i.e. 
\begin{equation}
\int\!\! d^D\!y\,G^{abc}_\nu(x,y)\,D^{\nu\nu'}_{cc'}(y,y') \sim
y_\nu D^{\nu\lambda}_{cc'}(y,y')  = 0  \,.
\end{equation}
This proves that the $g^2$ contribution to the gluon self energy indeed 
satisfies (\ref{tracon}). Again this demonstrates the consistency of our set
of radial gauge Feynman rules as well as the usefulness of the diagrammatic
rules presented in Sec.~\ref{sec:toe}.

\section{Conclusions} \label{sec:sum}

In this article we have presented a complete set of radial gauge Feynman rules
for non-Abelian gauge fields. The rather surprising results concerning 
the radial gauge field propagator have already been discussed in
\cite{leuwei96}:
\begin{itemize}
\item The propagator can be expressed in terms of a gauge invariant Wilson 
loop;
\item it is singular in four dimensional space;
\item this singularity is necessary to reproduce correct results for gauge
invariant quantities. 
\end{itemize}
The other properties of this set of rules are: 
\begin{itemize}
\item Feynman rules must be formulated in coordinate space since the gauge 
condition explicitly breaks translational invariance;
\item ghost fields decouple;
\item besides the conventional three and four gluon vertices no additional 
vertices are necessary. 
\end{itemize}
Introducing such a new set of rules it is
crucial to show its consistency with the conventional covariant gauge Feynman
rules, i.e.~to prove that results for gauge invariant
quantities calculated with both sets of rules agree with each other. To 
achieve that we have generalized the theorem of equivalence to coordinate
space Feynman rules. Originally it was developed by Cheng and Tsai to check
the consistency of different sets of Feynman rules formulated in momentum 
space. 

To the best of our knowledge such a systematic study of coordinate space 
Feynman rules has never been presented before. In \cite{delbourgo} the 
consistency of radial gauge Feynman rules was checked by an explicit one loop
calculation using basically the same technique as presented here. However it
was not pointed out there how this technique could be generalized to 
other Feynman diagrams involving higher order loops. In addition the radial 
gauge propagator used in \cite{delbourgo} was later shown to be incorrect 
\cite{moda}. 

A few additional remarks concerning the singularity of the radial gauge 
propagator (\ref{freeradprop}) are necessary: In principle there are three 
sources for a possible
singular behaviour. First, the already mentioned cusp singularity discussed
in great detail in \cite{leuwei96}. Second, an additional singularity shows up
if $x$ and $y$ are aligned with respect to the origin. These divergences are
already present in Euclidean space. In Minkowski space a third type of 
singularity appears if $x-y$ is light-like. Of course it would be interesting
to study also the latter two types of singularities in more detail. At least 
the third one is also connected to Wilson loops with part of the contour on 
the light cone \cite{korkor}. Despite the lack of a more detailed 
examination of these additional singularities we 
note that all these singularities are regularized by dimensional 
regularization. Thus also these divergences do not spoil the arguments 
presented in this article. 

To set up perturbation theory in radial gauge a complete set of Feynman rules 
is of course only the first step. Next one must figure out how renormalization
works in this gauge. Due to the lack of translational invariance one might
suspect that local counter terms are insufficient to renormalize the theory.
However some arguments that this is not the case are given in \cite{kumm}. 
Further general considerations as well as explicit loop calculations of 
self energies and vertex corrections are necessary to work out a 
renormalization scheme for radial gauge. We expect that the diagrammatic scheme
presented here might serve to simplify such calculations. 

\acknowledgements
I thank Ulrich Heinz for stimulating discussions and support. I also 
acknowledge useful discussions with Heribert Weigert and Marc Achhammer. 
This research was
supported in part by Deutsche Forschungsgemeinschaft and
Bundesministerium f\"ur Bildung, Wissenschaft, Forschung und
Technologie.

\begin{appendix}
\section{Feynman rules in coordinate space} \label{anh:feyn} 

Since Feynman rules are commonly used in momentum space we have collected 
some coordinate space expressions for propagators and vertices in this 
Appendix.

In Minkowski space with one time and $D-1$ space dimensions the propagator of 
a scalar particle is given by \cite{chengliseit}
\begin{eqnarray}
\Delta_S(x-y) &=& 
\int\! {d^D k \over (2\pi)^D} \, {-i\over k^2+i\eta} \, e^{-ik\cdot (x-y)} 
\nonumber \\
&=&  {\Gamma(D/2-1) \over 4\pi^{D/2}}\, [(x-y)^2-i\epsilon]^{1-D/2}  
\quad,\quad \eta,\epsilon \to +0  \,.  \label{skalarmink}
\end{eqnarray}
The Feynman propagator got additional factors of unity in Lorentz and color
space 
\begin{equation}
\langle B_\mu^a(x) B_\nu^b(y) \rangle_{\rm Feyn} 
=  \delta^{ab}\, D^F_{\mu\nu}(x,y) 
=   \delta^{ab}\, g_{\mu\nu} \,\Delta_S(x-y) \,.  \label{Feynprop}
\end{equation}
We define the (full) three gluon vertex ${\bf T}$ as 
\begin{eqnarray}
\lefteqn{\langle A_\lambda^a(w) \, A_\mu^b(x) \, A_\nu^c(y) \rangle =} 
\nonumber \\ &&  
\int\!\! d^D\!w' \,d^D\!x'\,d^D\!y'\, 
\langle A_\lambda^a(w) \, A_{\lambda'}^{a'}(w') \rangle
\langle A_\mu^b(x) \, A_{\mu'}^{b'}(x') \rangle
\langle A_\nu^c(y) \, A_{\nu'}^{c'}(y') \rangle
{\bf T}^{\lambda' \mu' \nu'}_{a'b'c'}(w',x',y')  \,;  
\end{eqnarray}
the four gluon vertex ${\bf Q}$ is defined in a completely analogous way. 

The expressions in lowest order of the coupling constant (bare vertices) are
given by 
\begin{eqnarray}
T^{\lambda \mu \nu}_{abc}(w,x,y) 
&=& -igf_{abc} \left[
  g^{\lambda\mu}(\partial^\nu_w - \partial^\nu_x)  \,  \delta(x-y)\,\delta(w-y)
\right. \nonumber\\&& \phantom{mmmn} 
+ g^{\mu\nu}(\partial^\lambda_x - \partial^\lambda_y)\,\delta(y-w)\,\delta(x-w)
\nonumber\\&& \left.\phantom{mmmn} 
+ g^{\nu\lambda}(\partial^\mu_y - \partial^\mu_w)  \,  \delta(w-x)\,\delta(y-x)
\right] \label{dreigluon}
\end{eqnarray}
and
\begin{eqnarray}
Q^{\kappa\lambda\mu\nu}_{abcd}(v,w,x,y)
&=& -i g^2 \left[
f_{abe} f_{cde}  (g^{\kappa\mu} g^{\nu\lambda} - g^{\kappa\nu} g^{\lambda\mu})
\right.\nonumber \\ && \!\!\phantom{mmm} 
+ f_{ace}f_{dbe} (g^{\kappa\nu} g^{\lambda\mu} - g^{\kappa\lambda} g^{\mu\nu})
\nonumber\\&& \!\!\phantom{mmm} \left.
+ f_{ade}f_{bce} (g^{\kappa\lambda} g^{\mu\nu} - g^{\kappa\mu} g^{\nu\lambda})
\right] \delta(v-w)\,\delta(v-x)\,\delta(v-y)  \,. \label{viergluon}
\end{eqnarray}

\section{Additional graphical rules} \label{rechen} 

Here we complete the set of diagrammatic rules presented in Sec.~\ref{sec:toe}.
In addition to 
Fig.~\ref{pardrei}, \ref{delstd}, \ref{einser}, and \ref{pfeilvier} we have 
to discuss the case that a contraction arrow points at a four gluon vertex
(Fig.~\ref{nurvier}) or at a ghost gluon vertex (Fig.~\ref{pfeileins}). 
The corresponding formulae are 
\begin{eqnarray}
\lefteqn{\int\!\! d^D\!v'\, 
g f_{eah} g^{\kappa}_{\phantom{\kappa}\kappa'} \delta(u-v)\,\delta(u-v')\,
Q^{\kappa'\lambda\mu\nu}_{hbcd}(v',w,x,y)} \nonumber \\ && {}
+ \int\!\! d^D\!w'\, 
g f_{ebh} g^{\lambda}_{\phantom{\lambda}\lambda'} \delta(u-w)\,\delta(u-w')\,
Q^{\kappa\lambda'\mu\nu}_{ahcd}(v,w',x,y)
\nonumber \\
&=& \int\!\! d^D\!x'\, 
g f_{ehc} g^{\mu}_{\phantom{\mu}\mu'} \delta(u-x)\,\delta(u-x')\,
Q^{\kappa\lambda\mu'\nu}_{abhd}(v,w,x',y)
\nonumber \\ && {}
+ \int\!\! d^D\!y'\, 
g f_{ehd} g^{\nu}_{\phantom{\nu}\nu'} \delta(u-y)\,\delta(u-y')\,
Q^{\kappa\lambda\mu\nu'}_{abch}(v,w,x,y')   
\end{eqnarray}
for Fig.~\ref{nurvier} and 
\begin{eqnarray}
\lefteqn{\int\!\!d^D\!v'\, G_\mu^{abe}(w,v')\, \delta(v-v')\, 
g f^{ecd}\delta(x-v')\,\delta(x-y) =} \nonumber \\ &&
\int\!\! d^D\!v'\, G_\nu^{dbe}(w,v')\,\delta(y-v')\,
g f^{eca} g^\nu_{\phantom{\nu}\mu} \delta(x-v) \, \delta(x-v')
\nonumber \\ && {}
+ \int\!\! d^D\!v'\, G_\nu^{ebc}(w,x)\,\delta(v'-x) \,
g f^{eda} g^\nu_{\phantom{\nu}\mu} \delta(y-v) \, \delta(y-v')  \,.
\end{eqnarray}
for Fig.~\ref{pfeileins}. 
The last relation is a somewhat complicated way to write down the Bianchi
identity for the structure constants.\footnote{$f^{abe}f^{cde}+ 
f^{ace}f^{dbe} + f^{ade}f^{bce}=0$.} It is missing in \cite{mitcheng}.

For Wilson loops we have to encounter the case that a partial derivative
(arrow) ``meets'' a contour integral, e.g. 
\begin{eqnarray}
\lefteqn{\int\limits_0^1 \!\! d\sigma_1 \int\limits_0^1 \!\! d\sigma_2
\int\limits_0^1 \!\! d\sigma_3 \,\Theta(\sigma_1 > \sigma_2 > \sigma_3) \,
\dot w_{\mu_1}(\sigma_1)\,  \dot w_{\mu_2}(\sigma_2)\, 
\dot w_{\mu_3}(\sigma_3)\,}
\nonumber\\&& \phantom{t}\times 
\int\!\! d^D\! x_1 \int\!\! d^D\! x_2 \int\!\! d^D\! x_3 \,
D^{\mu_1\nu_1}(w(\sigma_1),x_1) 
\partial^{\mu_2}_{w(\sigma_2)} \,\Delta^{\nu_2}(w(\sigma_2),x_2)
D^{\mu_3\nu_3}(w(\sigma_3),x_3) 
\nonumber\\&& \phantom{t}\times 
T^{a_1 a_2 a_3}_{\nu_1 \nu_2 \nu_3}(x_1,x_2,x_3)  \,
\mbox{Sp}(t_{a_1} t_{a_2} t_{a_3})
\nonumber\\&&
=  \int\limits_0^1 \!\! d\sigma_1 \int\limits_0^{\sigma_1} \!\! d\sigma_3
\int\limits_{\sigma_3}^{\sigma_1} \!\! d\sigma_2 \,
\dot w_{\mu_1}(\sigma_1)\,  \dot w_{\mu_2}(\sigma_2)\, 
\dot w_{\mu_3}(\sigma_3)\,
\nonumber\\&& \phantom{t}\times 
\int\!\! d^D\! x_1 \int\!\! d^D\! x_2 \int\!\! d^D\! x_3 \,
D^{\mu_1\nu_1}(w(\sigma_1),x_1) 
\partial^{\mu_2}_{w(\sigma_2)} \,\Delta^{\nu_2}(w(\sigma_2),x_2)
D^{\mu_3\nu_3}(w(\sigma_3),x_3) 
\nonumber\\&& \phantom{t}\times 
T^{a_1 a_2 a_3}_{\nu_1 \nu_2 \nu_3}(x_1,x_2,x_3)  \,
\mbox{Sp}(t_{a_1} t_{a_2} t_{a_3})  \,.
\end{eqnarray}
Here $w(\sigma)$, $\sigma \in [0,1]$, parameterizes an arbitrary Wilson loop
contour. We make use of the relation 
\begin{equation}
\dot w_{\mu_2} \partial^{\mu_2}_{w(\sigma_2)} = {d\over d\sigma_2}
\end{equation}
to evaluate the contour integral: 
\begin{eqnarray}
\lefteqn{\int\limits_0^1 \!\! d\sigma_1 
\int\limits_0^{\sigma_1} \!\! d\sigma_3  \,
\dot w_{\mu_1}(\sigma_1)\, \dot w_{\mu_3}(\sigma_3)\,
\int\!\! d^D\! x_1 \int\!\! d^D\! x_2 \int\!\! d^D\! x_3 \,
D^{\mu_1\nu_1}(w(\sigma_1),x_1) }
\nonumber\\&& \times 
\left[ \Delta^{\nu_2}(w(\sigma_1),x_2) 
- \Delta^{\nu_2}(w(\sigma_3),x_2) \right]
D^{\mu_3\nu_3}(w(\sigma_3),x_3) 
\nonumber\\&& \times 
T^{a_1 a_2 a_3}_{\nu_1 \nu_2 \nu_3}(x_1,x_2,x_3)  \,
\mbox{Sp}(t_{a_1} t_{a_2} t_{a_3})  \,.
\end{eqnarray}
This is depicted in Fig.~\ref{wilo}. In addition it is shown how the result 
is related to other diagrams \cite{chengwil}. 

\end{appendix}

\begin{figure}[ht]
\caption{The Wilson loop $W_1$ which is connected to the radial gauge field 
propagator.} \label{faecher}
\end{figure}
\begin{figure}[ht]
\caption{Partial derivative acting on a three gluon vertex.} \label{pardrei}
\end{figure}
\begin{figure}[ht]
\caption{Two loop vacuum diagrams.} \label{bubbleall}
\end{figure}
\begin{figure}[ht]
\caption{The sum of vacuum diagrams which ought to be gauge invariant.}
\label{bubblesym}
\end{figure}
\begin{figure}[ht]
\caption{Graphical decomposition of a propagator.} \label{propzer}
\end{figure}
\begin{figure}[ht]
\caption{Application of the diagrammatic rule shown in 
Fig.~\protect\ref{pardrei}.} 
\label{erster}
\end{figure}
\begin{figure}[ht]
\caption{How the partial derivative acts on a three gluon vertex with 
$\Delta_\mu$ on one leg.} 
\label{delstd}
\end{figure}
\begin{figure}[ht]
\caption{Application of the graphical relation shown in 
Fig.~\protect\ref{delstd}.} \label{zweiter}
\end{figure}
\begin{figure}[ht]
\caption{Product rule of differentiation at a ghost gluon vertex.} 
\label{einser}
\end{figure}
\begin{figure}[ht]
\caption{How to pull an arrow out of a closed ghost loop.}
\label{pull} 
\end{figure}
\begin{figure}[ht]
\caption{Calculation of all $G$-contributions from 
Fig.~\protect\ref{bubblesym} A.} 
\label{vierter}
\end{figure}
\begin{figure}[ht]
\caption{Decomposition of a propagator in a four gluon vertex diagram.} 
\label{vierzer}
\end{figure}
\begin{figure}[ht]
\caption{A relation connecting three and four gluon vertices.} 
\label{pfeilvier}
\end{figure}
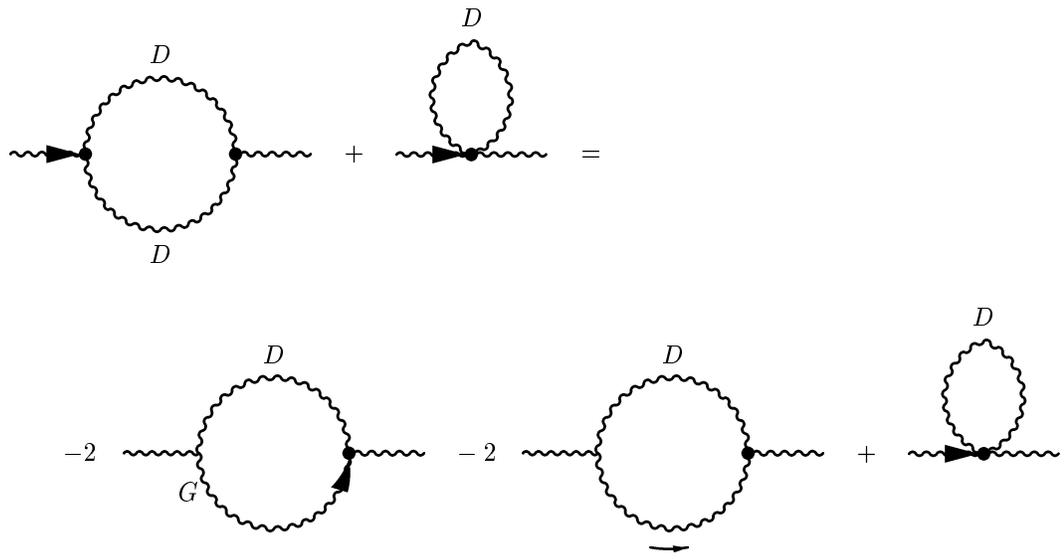
\begin{figure}[ht]
\caption{Transversality of the one loop self energy contribution.} 
\label{trglse}
\end{figure}
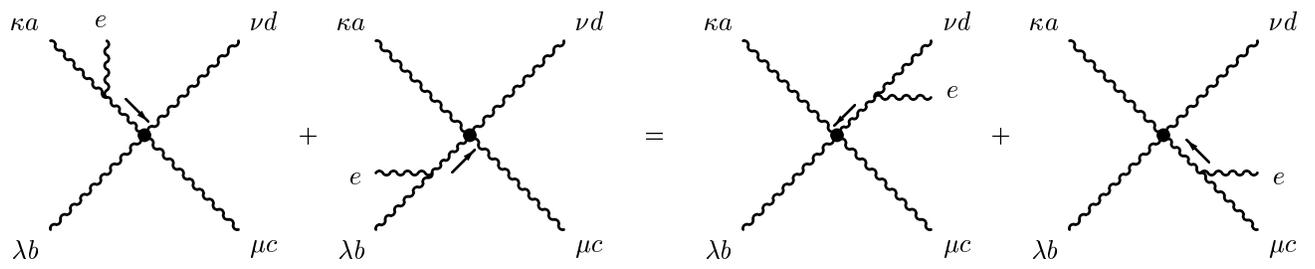
\begin{figure}[ht]
\caption{A relation connecting four gluon vertices.} \label{nurvier}
\end{figure}
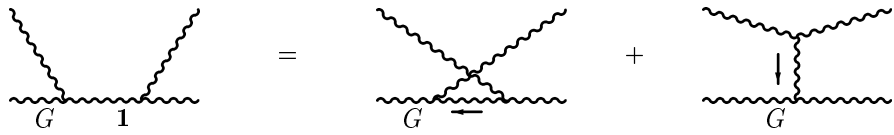
\begin{figure}[ht]
\caption{A relation connecting ghost vertices.} \label{pfeileins}
\end{figure}
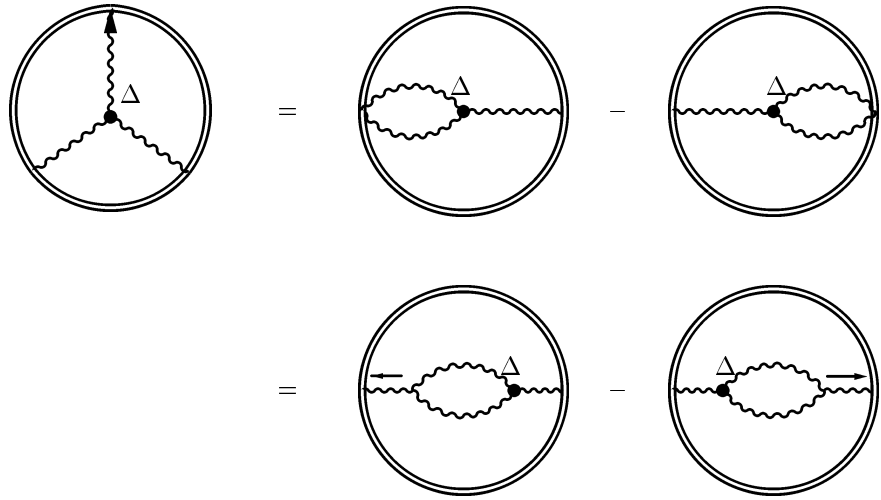
\begin{figure}[ht]
\caption{Additional relation for Wilson loops.} \label{wilo}
\end{figure}

\end{document}